\documentclass[RNAAS]{aastex62}

\newcommand\gaia{\textit{Gaia}}
\newcommand\gdr[1]{\gaia~DR#1}

\begin{document}

\title{Pre-main sequence population of Sco-Cen unveiled with Gaia DR2}

\correspondingauthor{Jorge A.~Villa V\'{e}lez}
\email{jvilla@strw.leidenuniv.nl}

\author[0000-0002-8300-3754]{Jorge A.~Villa V\'{e}lez}
\affiliation{Leiden Observatory, Leiden University, P.O. Box 9513, 2300 RA Leiden, The Netherlands}

\author[0000-0002-7419-9679]{Anthony G.A.~Brown}
\affiliation{Leiden Observatory, Leiden University, P.O. Box 9513, 2300 RA Leiden, The Netherlands}

\author[0000-0002-7064-8270]{Matthew A.~Kenworthy}
\affiliation{Leiden Observatory, Leiden University, P.O. Box 9513, 2300 RA Leiden, The Netherlands}

%% Note that RNAAS manuscripts DO NOT have abstracts.
%% See the online documentation for the full list of available subject
%% keywords and the rules for their use.

%% Start the main body of the article. If no sections in the 
%% research note leave the \section call blank to make the title.
\section{} 

The characterization of the stellar content of OB associations has traditionally relied on kinematic information from proper motions and radial velocities to separate the association members from the foreground and background field star population \citep{deZeeuw99, Wright18}. This led to a bias in the association membership toward the more massive stars, as kinematic information was usually lacking for the low-mass pre-main sequence members. 
Dedicated spectroscopic and photometric surveys uncovered only a limited fraction of the pre-main sequence population in, e.g., the Sco OB2 and Orion OB1 associations \citep[e.g.][]{Preibisch02, Preibisch08, Briceno18, Kounkel17}. 
%Spectroscopic surveys are very time consuming, while the photometric surveys suffered from a lack of accurate luminosity information to separate the pre-main sequence from the main sequence population. 

\begin{figure}
\begin{center}
\includegraphics[scale=0.5,angle=0]{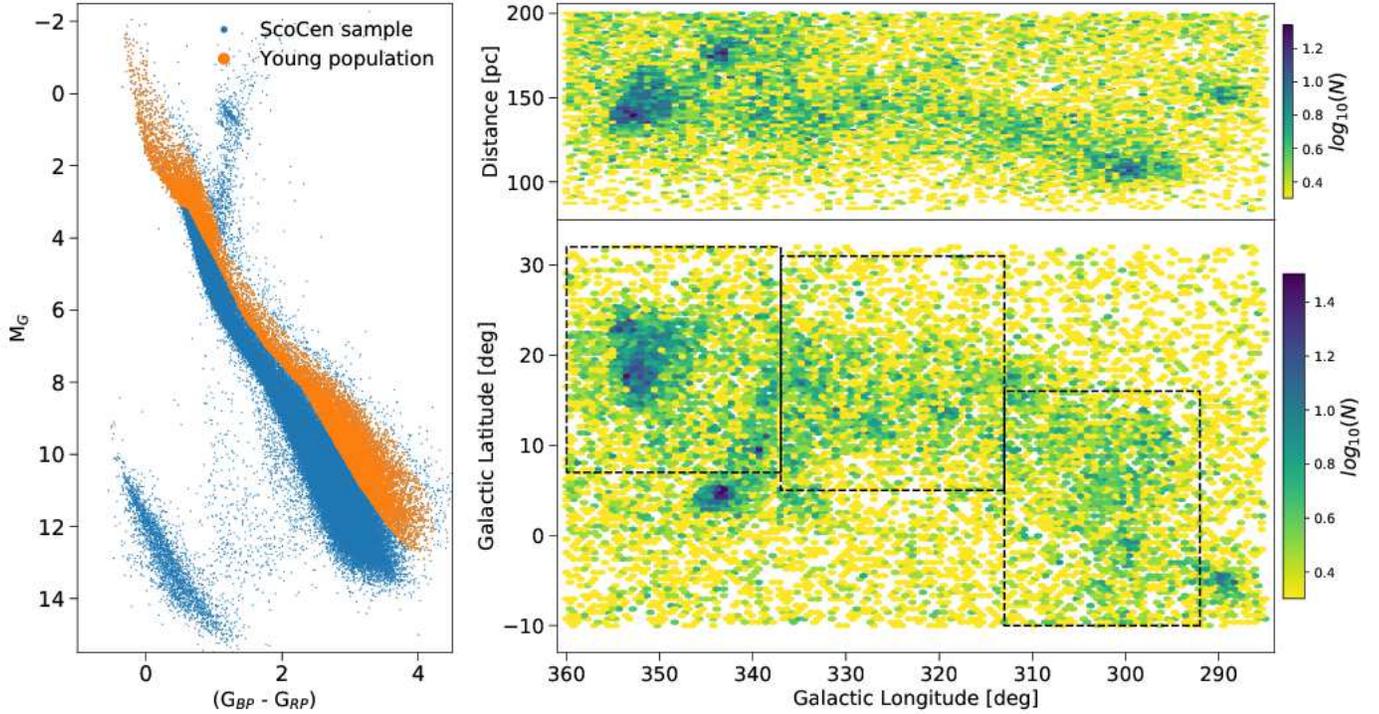}
\caption{\textbf{Left:} Color-magnitude diagram of the Sco-Cen region with the young stellar population highlighted in orange. \textbf{Bottom right:} Sky distribution of the young stellar population. The three subgroups of Sco OB2 (from left to right, Upper Scorpius, Upper Centaurus Lupus, and Lower Centaurus Crux), as defined in \cite{deZeeuw99}, are indicated. \textbf{Top right:} Distance distribution (with distance calculated as $1/\varpi$) as a function of Galactic longitude.\label{fig:cmd}}
\end{center}
\end{figure}

\cite{Zari17} used the \gdr{1} \citep{Prusti16,Brown16} data in combination with 2MASS photometry \citep{Skrutskie06} to isolate the pre-main sequence population in the apparent magnitude vs.\ color diagram and map the distribution of the young stars in the Orion region.
%%% \cite{Zari17} noticed that stars with small proper motions in the parallax slice $2\leq\varpi\leq3.5$~mas show a marked concentration on the sky. They subsequently used the color apparent magnitude diagram for all \gdr{1} sources in the Orion region for which 2MASS photometry was available to isolate the pre-main sequence population located at $\varpi\sim2.65$~mas. The identification of this population was relatively straightforward because the large distance to the Orion population implies less impact of the extent of the association along the line of sight on the spread in apparent magnitudes. 
\cite{Kounkel18} subsequently used \gdr{2} \citep{Brown18} parallaxes and photometry to construct an observational Hertzsprung-Russell diagram in which they more precisely isolated the pre-main sequence population in Orion.
We apply this technique to the Scorpius-Lupus-Centaurus-Crux area on the sky, which contains the Sco OB2 association \citep{Blaauw46}. We select from \gdr{2} all stars with galactic coordinates $285^\circ\leq\ell\leq360^\circ$ and $-10^\circ\leq b\leq+32^\circ$, thus covering a wider area around the traditional boundaries of the association (see Figure~\ref{fig:cmd}), and with parallaxes between $5$ and $12$~mas, covering the known distances to the Sco OB2 association subgroups \citep{Wright18}. The relative parallax errors were restricted to $<10\%$ and sources with potentially spurious parallax values or poor photometry \citep[Appendix C of][]{Lindegren18} were removed. The resulting sample contains $120\,911$ sources for which the observational HR diagram is shown in the left panel of Figure~\ref{fig:cmd}. The pre-main sequence population is clearly separated from the main sequence at colors $(G_\mathrm{BP}-G_\mathrm{RP})>1$, where at $(G_\mathrm{BP}-G_\mathrm{RP})>2$ the separation is well above the $0.75$ magnitude expected from a population of equal mass binaries. We proceeded to isolate this population as well as the young early type stars through a selection by hand in the color-absolute magnitude space (the selection polygon is available from Villa V\'{e}lez et al. 2018, Data\_Selection.ipynb, v1.0.0, Zenodo, \href{https://zenodo.org/record/1286576#.Wx6NAmMzaV4}{doi:10.5281/zenodo.1286576}). The distribution of the $14\,459$ selected stars on the sky is shown in the right panel of Figure~\ref{fig:cmd}. 

There is a very clear concentration of the young stellar population which follows the traditional boundaries of the Sco OB2 association, consistent with most of the selected sources being association members. The Upper Scorpius region stands out as the densest concentration of young stars with the sparser distribution in the Upper Centaurus Lupus and Lower Centaurus Crux areas showing clear hints of clumps of young stars \citep[indications of substructure were also found by][]{deZeeuw99}. The concentration of sources near $(\ell,b)=(290^\circ,-5^\circ)$ corresponds to the IC 2602 cluster \citep[$\varpi=6.74\pm0.25$~mas;][]{vanLeeuwen17}. Our expanded search reveals an additional population of young stars potentially associated with Sco OB2 ($b\sim5^\circ$ and $\ell\sim345^\circ$) at a mean distance of $\sim180\mathrm{pc}$ ($5$-$6$~mas). This population was also noted by \citet{deZeeuw99} (their Section 4.5 and Figure 9), and in \citet{Mamajek2016}.\\

%Figure~\ref{fig:cmd} illustrates the potential for using \gdr{2} to identify the pre-main sequence population in the vicinity of the Sun through an inspection of the observational HR diagram. The resulting maps of the pre-main sequence population can be used to guide more detailed membership studies of OB associations, including a thorough re-assessment of the traditional association and subgroup boundaries. We anticipate a major revision in our knowledge of OB associations coupled to new insights into the triggering and propagation of star formation.

%The software is available at: \url{https://github.com/Jurgenvilla/Gaia_DR2_ScoCen}

%The software is available at Villa V\'{e}lez et al. 2018, Data\_Selection.ipynb, v1.0.0, Zenodo, \href{https://zenodo.org/record/1286576#.Wx6NAmMzaV4}{doi:10.5281/zenodo.1286576}, as developed on \href{https://github.com/Jurgenvilla/Gaia_DR2_ScoCen/releases}{GitHub}.

\acknowledgments

We used data from the European Space Agency (ESA) mission {\gaia} (\url{https://www.cosmos.esa.int/gaia}), processed by the {\gaia} Data Processing and Analysis Consortium (DPAC, \url{https://www.cosmos.esa.int/web/gaia/dpac/consortium}). Funding for the DPAC is provided by national institutions, in particular those participating in the {\gaia} Multilateral Agreement. This research made use of Astropy\footnote{\url{http://www.astropy.org/}}, a community-developed core Python package for Astronomy \citep{Robitaille13}, and matplotlib\footnote{\url{http://matplotlib.org/}} for plotting the figures \citep{Hunter07}.

\bibliography{report}
\end{document}